\pdfoutput=1

\documentclass[12pt,a4paper]{article}

\usepackage{ifthen} 
\newboolean{pdflatex}
\setboolean{pdflatex}{true} 

\newboolean{articletitles}
\setboolean{articletitles}{true} 

\newboolean{uprightparticles}
\setboolean{uprightparticles}{false} 

\newboolean{inbibliography}
\setboolean{inbibliography}{false} 

\usepackage{microtype}
\usepackage{lineno}  
\usepackage{xspace} 
\usepackage{caption} 

\usepackage{graphicx}  
\usepackage{color}
\usepackage{colortbl}
\graphicspath{{./figs/}} 

\usepackage{amsmath} 
\usepackage{amssymb}
\usepackage{amsfonts}
\usepackage{upgreek} 

\newcommand*\patchAmsMathEnvironmentForLineno[1]{%
\expandafter\let\csname old#1\expandafter\endcsname\csname #1\endcsname
\expandafter\let\csname oldend#1\expandafter\endcsname\csname
end#1\endcsname
 \renewenvironment{#1}%
   {\linenomath\csname old#1\endcsname}%
   {\csname oldend#1\endcsname\endlinenomath}%
}
\newcommand*\patchBothAmsMathEnvironmentsForLineno[1]{%
  \patchAmsMathEnvironmentForLineno{#1}%
  \patchAmsMathEnvironmentForLineno{#1*}%
}
\AtBeginDocument{%
\patchBothAmsMathEnvironmentsForLineno{equation}%
\patchBothAmsMathEnvironmentsForLineno{align}%
\patchBothAmsMathEnvironmentsForLineno{flalign}%
\patchBothAmsMathEnvironmentsForLineno{alignat}%
\patchBothAmsMathEnvironmentsForLineno{gather}%
\patchBothAmsMathEnvironmentsForLineno{multline}%
\patchBothAmsMathEnvironmentsForLineno{eqnarray}%
}

\usepackage{hyperref}    
\usepackage[all]{hypcap} 

\def\lhcb {\mbox{LHCb}\xspace}

\def\babar  {\mbox{BaBar}\xspace}
\def\belle  {\mbox{Belle}\xspace}
\def\cleo   {\mbox{CLEO}\xspace}

\def\lhc    {\mbox{LHC}\xspace}

\ifthenelse{\boolean{uprightparticles}}%
{

 \def\Peta        {\ensuremath{\upeta}\xspace}

 \def\Ppi         {\ensuremath{\uppi}\xspace}

 \def\Pchi        {\ensuremath{\upchi}\xspace}                 
 \def\Ppsi        {\ensuremath{\uppsi}\xspace}

 \def\PDelta      {\ensuremath{\Delta}\xspace}                 
 \def\PXi      {\ensuremath{\Xi}\xspace}                 
 \def\PLambda      {\ensuremath{\Lambda}\xspace}                 
 \def\PSigma      {\ensuremath{\Sigma}\xspace}                 
 \def\POmega      {\ensuremath{\Omega}\xspace}                 
 \def\PUpsilon      {\ensuremath{\Upsilon}\xspace}                 
 
 \def\PB      {\ensuremath{\mathrm{B}}\xspace}                 
                  
 \def\PD      {\ensuremath{\mathrm{D}}\xspace}

 \def\PJ      {\ensuremath{\mathrm{J}}\xspace}                 
 \def\PK      {\ensuremath{\mathrm{K}}\xspace}

 \def\Pb      {\ensuremath{\mathrm{b}}\xspace}                 
 \def\Pc      {\ensuremath{\mathrm{c}}\xspace}

 \def\Pi      {\ensuremath{\mathrm{i}}\xspace}

 \def\Pp      {\ensuremath{\mathrm{p}}\xspace}

 \def\Ps      {\ensuremath{\mathrm{s}}\xspace}

}
{

 \def\Peta        {\ensuremath{\eta}\xspace}

 \def\Ppi         {\ensuremath{\pi}\xspace}

 \def\Pchi        {\ensuremath{\chi}\xspace}                 
 \def\Ppsi        {\ensuremath{\psi}\xspace}                 
                  
 \mathchardef\PDelta="7101
 \mathchardef\PXi="7104
 \mathchardef\PLambda="7103
 \mathchardef\PSigma="7106
 \mathchardef\POmega="710A
 \mathchardef\PUpsilon="7107
                  
 \def\PB      {\ensuremath{B}\xspace}                 
                  
 \def\PD      {\ensuremath{D}\xspace}

 \def\PJ      {\ensuremath{J}\xspace}                 
 \def\PK      {\ensuremath{K}\xspace}

 \def\Pb      {\ensuremath{b}\xspace}                 
 \def\Pc      {\ensuremath{c}\xspace}

 \def\Pi      {\ensuremath{i}\xspace}

 \def\Pp      {\ensuremath{p}\xspace}

 \def\Ps      {\ensuremath{s}\xspace}

}

\makeatletter
\ifcase \@ptsize \relax
  \newcommand{\miniscule}{\@setfontsize\miniscule{4}{5}}
\or
  \newcommand{\miniscule}{\@setfontsize\miniscule{5}{6}}
\or
  \newcommand{\miniscule}{\@setfontsize\miniscule{5}{6}}
\fi
\makeatother

\DeclareRobustCommand{\optbar}[1]{\shortstack{{\miniscule (\rule[.5ex]{1.25em}{.18mm})}
  \\ [-.7ex] $#1$}}

\def\squark    {{\ensuremath{\Ps}}\xspace}

\def\cquark    {{\ensuremath{\Pc}}\xspace}

\def\bquark    {{\ensuremath{\Pb}}\xspace}

\def\pion   {{\ensuremath{\Ppi}}\xspace}
\def\piz    {{\ensuremath{\pion^0}}\xspace}

  \def\Kbar    {{\kern 0.2em\overline{\kern -0.2em \PK}{}}\xspace}

\def\KorKbar    {\kern 0.18em\optbar{\kern -0.18em K}{}\xspace}

  \def\Dbar    {{\kern 0.2em\overline{\kern -0.2em \PD}{}}\xspace}

\def\DorDbar    {\kern 0.18em\optbar{\kern -0.18em D}{}\xspace}

\def\B       {{\ensuremath{\PB}}\xspace}
\def\Bbar    {{\ensuremath{\kern 0.18em\overline{\kern -0.18em \PB}{}}}\xspace}

\def\BorBbar    {\kern 0.18em\optbar{\kern -0.18em B}{}\xspace}

\def\Bzb     {{\ensuremath{\Bbar{}^0}}\xspace}

\def\Bub     {{\ensuremath{\B^-}}\xspace}

\def\Bsb     {{\ensuremath{\Bbar{}^0_\squark}}\xspace}

\def\Bcm     {{\ensuremath{\B_\cquark^-}}\xspace}

\def\jpsi     {{\ensuremath{{\PJ\mskip -3mu/\mskip -2mu\Ppsi\mskip 2mu}}}\xspace}
\def\psitwos  {{\ensuremath{\Ppsi{(2S)}}}\xspace}

\def\etac     {{\ensuremath{\Peta_\cquark}}\xspace}
\def\chiczero {{\ensuremath{\Pchi_{\cquark 0}}}\xspace}
\def\chicone  {{\ensuremath{\Pchi_{\cquark 1}}}\xspace}
\def\chictwo  {{\ensuremath{\Pchi_{\cquark 2}}}\xspace}
  \def\Y#1S{\ensuremath{\PUpsilon{(#1S)}}\xspace}

\def\chic  {{\ensuremath{\Pchi_{c}}}\xspace}

\def\proton      {{\ensuremath{\Pp}}\xspace}
\def\antiproton  {{\ensuremath{\overline \proton}}\xspace}

\def\Lbar        {{\ensuremath{\kern 0.1em\overline{\kern -0.1em\PLambda}}}\xspace}
\def\LorLbar    {\kern 0.18em\optbar{\kern -0.18em \PLambda}{}\xspace}

\def\BF         {{\ensuremath{\cal B}}\xspace}

\def\BR         {\BF}
\newcommand{\decay}[2]{\ensuremath{#1\!\to #2}\xspace}         

\def\to                 {\ensuremath{\rightarrow}\xspace}

\def\AT#1     {\ensuremath{A_{\mathrm{T}}^{#1}}\xspace}           

\def\C#1      {\ensuremath{\mathcal{C}_{#1}}\xspace}                       
\def\Cp#1     {\ensuremath{\mathcal{C}_{#1}^{'}}\xspace}                    
\def\Ceff#1   {\ensuremath{\mathcal{C}_{#1}^{\mathrm{(eff)}}}\xspace}        
\def\Cpeff#1  {\ensuremath{\mathcal{C}_{#1}^{'\mathrm{(eff)}}}\xspace}       
\def\Ope#1    {\ensuremath{\mathcal{O}_{#1}}\xspace}                       
\def\Opep#1   {\ensuremath{\mathcal{O}_{#1}^{'}}\xspace}                    

\newcommand{\tev}{\ifthenelse{\boolean{inbibliography}}{\ensuremath{~T\kern -0.05em eV}\xspace}{\ensuremath{\mathrm{\,Te\kern -0.1em V}}}\xspace}
\newcommand{\gev}{\ensuremath{\mathrm{\,Ge\kern -0.1em V}}\xspace}
\newcommand{\mev}{\ensuremath{\mathrm{\,Me\kern -0.1em V}}\xspace}
\newcommand{\kev}{\ensuremath{\mathrm{\,ke\kern -0.1em V}}\xspace}
\newcommand{\ev}{\ensuremath{\mathrm{\,e\kern -0.1em V}}\xspace}
\newcommand{\gevc}{\ensuremath{{\mathrm{\,Ge\kern -0.1em V\!/}c}}\xspace}
\newcommand{\mevc}{\ensuremath{{\mathrm{\,Me\kern -0.1em V\!/}c}}\xspace}
\newcommand{\gevcc}{\ensuremath{{\mathrm{\,Ge\kern -0.1em V\!/}c^2}}\xspace}
\newcommand{\gevgevcccc}{\ensuremath{{\mathrm{\,Ge\kern -0.1em V^2\!/}c^4}}\xspace}
\newcommand{\mevcc}{\ensuremath{{\mathrm{\,Me\kern -0.1em V\!/}c^2}}\xspace}

\def\mum  {\ensuremath{{\,\upmu\rm m}}\xspace}

\def\nb {\ensuremath{\rm \,nb}\xspace}

\def\fs   {\ensuremath{\rm \,fs}\xspace}

\def\gsim{{~\raise.15em\hbox{$>$}\kern-.85em
          \lower.35em\hbox{$\sim$}~}\xspace}
\def\lsim{{~\raise.15em\hbox{$<$}\kern-.85em
          \lower.35em\hbox{$\sim$}~}\xspace}

\def\sqs   {\ensuremath{\protect\sqrt{s}}\xspace}

\def\pt         {\mbox{$p_{\rm T}$}\xspace}

\def\evtgen     {\mbox{\textsc{EvtGen}}\xspace}

\def\geant      {\mbox{\textsc{Geant4}}\xspace}

\def\photos     {\mbox{\textsc{Photos}}\xspace}

\def\pythia     {\mbox{\textsc{Pythia}}\xspace}

\def\tell1  {TELL1\xspace}
\def\ukl1   {UKL1\xspace}

\usepackage{cite} 
\usepackage{mciteplus}

\usepackage{longtable} 

\begin{document}

\renewcommand{\thefootnote}{\fnsymbol{footnote}}
\setcounter{footnote}{1}


\begin{titlepage}
\pagenumbering{roman}

\vspace*{-1.5cm}
\centerline{\large EUROPEAN ORGANIZATION FOR NUCLEAR RESEARCH (CERN)}
\vspace*{1.5cm}
\hspace*{-0.5cm}
\begin{tabular*}{\linewidth}{lc@{\extracolsep{\fill}}r}
\ifthenelse{\boolean{pdflatex}}
{\vspace*{-2.7cm}\mbox{\!\!\!\includegraphics[width=.14\textwidth]{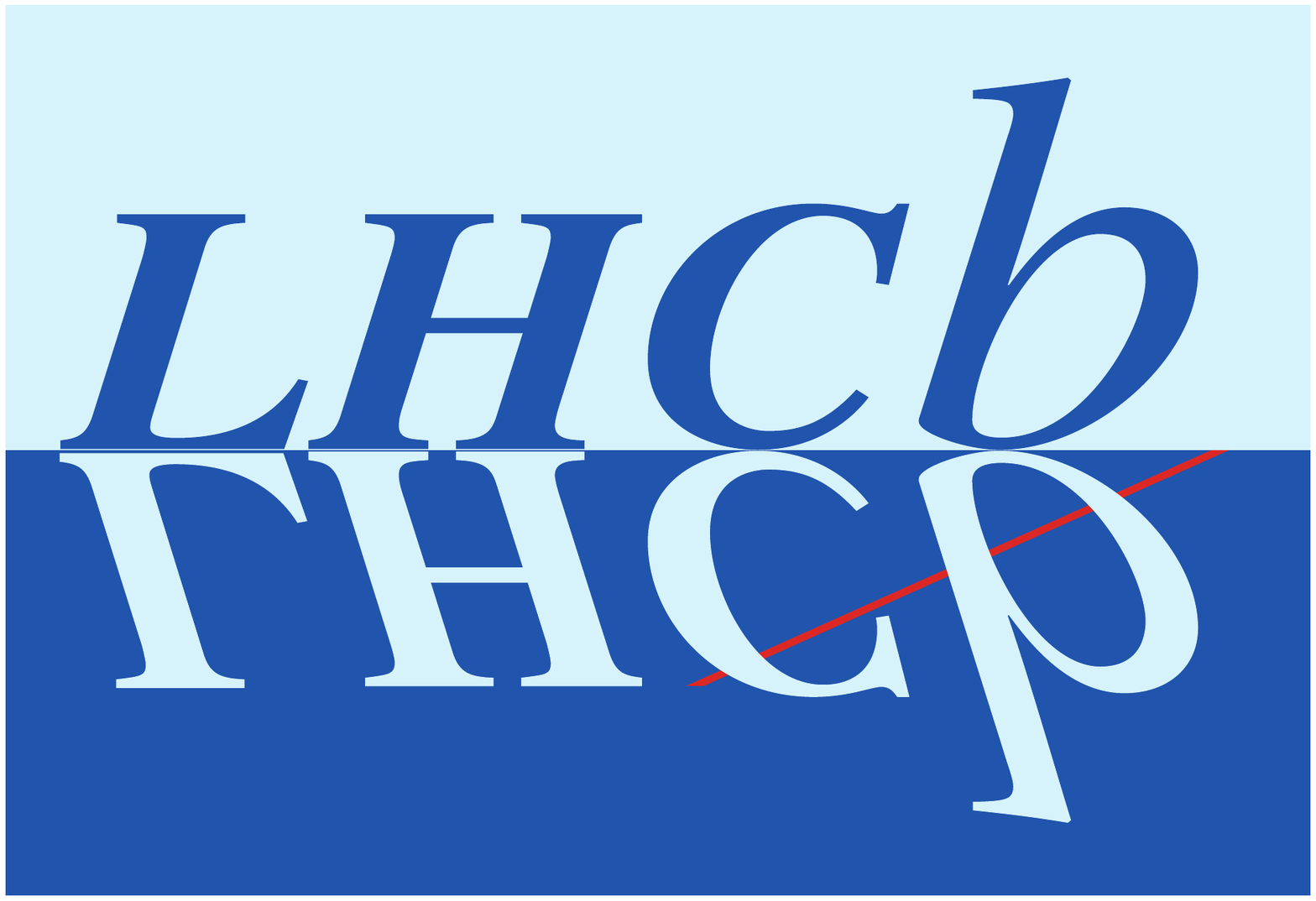}} & &}%
{\vspace*{-1.2cm}\mbox{\!\!\!\includegraphics[width=.12\textwidth]{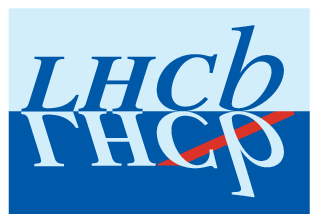}} & &}%
\\
 & & CERN-PH-EP-2014-218 \\  
 & & LHCb-PAPER-2014-029 \\  
 & & 11 September 2014 \\ 
 & & \\
\end{tabular*}

\vspace*{2.0cm}

{\bf\boldmath\huge
\begin{center}
Measurement of the $\etac(1S)$ production cross-section in proton-proton collisions via the decay \decay{\etac(1S)}{\proton \antiproton} 
\end{center}
}

\vspace*{0.8cm}

\begin{center}
The LHCb collaboration\footnote{Authors are listed at the end of this paper.}
\end{center}

\vspace{\fill}

\begin{abstract}
  \noindent
The production of the $\etac (1S)$ state in proton-proton collisions 
is probed via its decay to the \proton\antiproton final state with the \lhcb detector, 
in the rapidity range $2.0 < y < 4.5$ and in the meson transverse-momentum range $\pt > 6.5 \gevc$. 
The cross-section for prompt production of $\etac (1S)$ mesons 
relative to the prompt \jpsi cross-section
is measured, for the first time, to be 
$\sigma_{\etac (1S)}/\sigma_{\jpsi} = 1.74 \pm 0.29 \pm 0.28 \pm 0.18 _{\BF}$
at a centre-of-mass energy $\sqs = 7 \tev$ using data corresponding to an integrated luminosity of 0.7 fb$^{-1}$, 
and 
$\sigma_{\etac (1S)}/\sigma_{\jpsi} = 1.60 \pm 0.29 \pm 0.25 \pm 0.17 _{\BF}$
at 
$\sqs = 8 \tev$ using 
2.0 fb$^{-1}$. 
The uncertainties quoted are, in order, statistical, systematic, and that on the ratio of branching fractions 
of the $\etac (1S)$ and \jpsi decays to the \proton\antiproton final state. 
In addition, the inclusive branching fraction of \bquark-hadron decays into $\etac (1S)$ mesons is measured, for the first time, to be 
$\BR ( b \to \etac X ) = (4.88 \pm 0.64 \pm 0.29 \pm 0.67 _{\BF}) \times 10^{-3}$,
where the third uncertainty includes also the uncertainty on the \jpsi inclusive branching fraction from \bquark-hadron decays. 
The difference between the \jpsi and $\etac (1S)$ meson masses 
is determined to be $114.7 \pm 1.5 \pm 0.1 \mevcc$. 
  
\end{abstract}

\vspace*{0.2cm}

\begin{center}
  Submitted to Eur.~Phys.~J.~C
\end{center}

\vspace{\fill}

{\footnotesize 
\centerline{\copyright~CERN on behalf of the \lhcb collaboration, license \href{http://creativecommons.org/licenses/by/4.0/}{CC-BY-4.0}.}}
\vspace*{2mm}

\end{titlepage}


\newpage
\setcounter{page}{2}
\mbox{~}

\cleardoublepage


\renewcommand{\thefootnote}{\arabic{footnote}}
\setcounter{footnote}{0}



\pagestyle{plain} 
\setcounter{page}{1}
\pagenumbering{arabic}


%

\section{Introduction}
\label{sec:intro}

High centre-of-mass energies available in proton-proton collisions at the \lhc 
allow models describing charmonium production to be tested. 
We distinguish promptly produced charmonia from those originating from \bquark-hadron decays.
Promptly produced charmonia include charmonia directly produced in parton interactions 
and those originating from the decays of heavier quarkonium states, which are in turn produced in parton interactions. 
While measurements of \jpsi and \psitwos meson production rates at the 
\lhc~\cite{LHCb-PAPER-2011-003,LHCb-PAPER-2013-008,LHCb-PAPER-2013-016,Chatrchyan:2011kc,Aad:2011sp,Abelev:2012gx} 
are successfully described by next-to-leading order (NLO) calculations 
in non-relativistic quantum chromodynamics (QCD)~\cite{Ma:2010yw}, 
the observation of small or no polarization in \jpsi meson prompt
production~\cite{LHCb-PAPER-2013-008} 
remains unexplained within the available theoretical framework~\cite{Brambilla:2010cs}. 
The investigation of the lowest state, the $\etac (1S)$ meson, can provide important additional information
on the long-distance matrix elements~\cite{Butenschoen:2012px,Chao:2012iv}. 
In particular, the heavy-quark spin-symmetry relation between the $\etac (1S)$ and \jpsi matrix elements can be tested, 
with the NLO calculations predicting a different dependence of the production rates 
on charmonium transverse momentum, \pt, 
for spin singlet ($\etac (1S)$) and triplet (\jpsi, $\chi_{cJ}$) 
states~\cite{Maltoni:2004hv,Petrelli:1997ge,Kuhn:1992qw}. 
Thus, a measurement of the \pt dependence of the $\etac (1S)$ production rate, in particular in the low \pt region, 
can have important implications.
Recent \lhcb results on prompt production of \chic states~\cite{LHCb-PAPER-2013-028} 
provide information on the production of the $P$-wave states \chiczero and \chictwo 
at low \pt, using the well-understood \chicone production as a reference. 
A measurement of the cross-section of prompt $\etac (1S)$ production 
may allow an important comparison with the \chiczero results and 
yields indirect information on the production of heavier states. 

At \lhc energies, all \bquark-hadron species are produced, including weakly decaying 
\Bub, \Bzb, \Bsb, \Bcm mesons, \bquark-baryons, and their charge-conjugate states. 
A previous study of inclusive $\etac (1S)$ meson production in \bquark-hadron decays by the \cleo experiment, 
based on a sample of \Bub and \Bzb mesons, 
placed an upper limit on the combined inclusive branching fraction of \Bub and \Bzb meson decays 
into final states containing an $\etac (1S)$ meson of 
$\BF ( \Bub , \Bzb \to \etac (1S) X ) < 9 \times 10^{-3}$ at $90 \%$~confidence level~\cite{Balest:1994jf}.
Exclusive analyses of $\etac (1S)$ and \jpsi meson production in \bquark-hadron decays using 
the $\B \to K ( \proton \antiproton)$ decay mode 
have been performed by the \babar experiment~\cite{Aubert:2005gw}, 
by the \belle experiment~\cite{Wei:2007fg} and recently by the \lhcb experiment~\cite{LHCb-PAPER-2012-047}. 

In the present paper we report the first measurement of the cross-section for the prompt production 
of $\etac (1S)$ mesons 
in \proton\proton collisions at $\sqs = 7 \tev$ and $\sqs = 8 \tev$ centre-of-mass energies, 
as well as the \bquark-hadron inclusive branching fraction into $\etac (1S)$ final states. 
This paper extends the scope of previous charmonium production studies reported 
by \lhcb, which were restricted to the use of \jpsi or \psitwos decays to dimuon final 
states~\cite{LHCb-PAPER-2011-003,LHCb-PAPER-2013-008,LHCb-PAPER-2011-045,LHCb-PAPER-2013-028}. 
In order to explore states that do not have $J^{PC} = 1^{--}$ quantum numbers, 
while avoiding reconstruction of radiative decays with low-energy photons, 
the authors of Ref.~\cite{Barsuk:2012ic} suggested to investigate hadronic final states. 
In the present analysis, we reconstruct $\etac (1S)$ mesons decaying into the \proton\antiproton final state. 
All well-established charmonium states decay to \proton\antiproton final states~\cite{PDG2014,Barsuk:2012ic}.
With its powerful charged-hadron identification and high charmonium production rate, 
the \lhcb experiment is well positioned for these studies. 
The measurements are performed relative to the topologically and kinematically similar 
$\jpsi\to \proton\antiproton$ channel, 
which allows partial cancellation of systematic uncertainties in the ratio.
This is the first such inclusive analysis using decays to hadronic final states 
performed at a hadron collider. 

In addition, a departure in excess of two standard deviations 
between the recent BES~III results~\cite{BESIII:2011ab,Ablikim:2012ur} 
and earlier measurements~\cite{PDG2014} motivates the determination of 
the difference between \jpsi and $\etac (1S)$ meson masses $\Delta M _{\jpsi , \, \etac (1S)} \equiv M_{\jpsi} - M_{\etac (1S)}$ 
using a different technique and final state. 
In the present analysis, the low-background sample of charmonia produced in \bquark-hadron decays
is used to determine $\Delta M _{\jpsi , \, \etac (1S)}$ and the $\etac (1S)$ natural width, $\Gamma_{\etac (1S)}$. 

In section~\ref{sec:lhcb} we present the \lhcb detector and data sample used for the analysis. 
Section~\ref{sec:ana} describes the analysis details, while the systematic uncertainties 
are discussed in section~\ref{sec:syst}. 
The results are given in section~\ref{sec:results} and summarized in section~\ref{sec:summary}.

\section{\lhcb detector and data sample}
\label{sec:lhcb}

The \lhcb detector~\cite{Alves:2008zz} is a single-arm forward
spectrometer covering the \mbox{pseudorapidity} range $2< \eta <5$,
designed for the study of particles containing \bquark or \cquark
quarks. The detector includes a high-precision tracking system
consisting of a silicon-strip vertex detector surrounding the $pp$
interaction region, a large-area silicon-strip detector located
upstream of a dipole magnet with a bending power of about
$4{\rm\,T m}$, and three stations of silicon-strip detectors and straw
drift tubes placed downstream of the magnet.
The combined tracking system provides a momentum measurement with
a relative uncertainty that varies from 0.4\% at low momentum 
to 0.6\% at 100\gevc,
and an impact parameter measurement with a resolution of 20\mum for
charged particles with large transverse momentum. 
Different types of charged hadrons are distinguished using information
from two ring-imaging Cherenkov detectors. 
Photon, electron, and hadron candidates are identified by a system 
consisting of scintillating-pad and preshower detectors, an electromagnetic
calorimeter, and a hadronic calorimeter. 
Muons are identified by a system composed of alternating layers of iron and multiwire
proportional chambers.
The trigger consists of a hardware stage, based on information from the calorimeter 
and muon systems, followed by a software stage, which applies a full event
reconstruction.

Events enriched in signal decays are selected by the hardware trigger, 
based on the presence of a single high-energy deposit in the calorimeter.
The subsequent software trigger specifically rejects high-multiplicity events 
and selects events with two oppositely charged particles 
having good track-fit quality 
and transverse momentum larger than 1.9\gevc. 
Proton and antiproton candidates are 
identified using the information from Cherenkov and tracking detectors~\cite{LHCb-DP-2012-003}. 
Selected \proton and \antiproton candidates are required to form a good quality vertex. 
In order to further suppress the dominant background from accidental combinations 
of random tracks (combinatorial background),
charmonium candidates are required to have high transverse momentum, $\pt > 6.5 \gevc$. 

The present analysis uses \proton\proton collision data recorded by the \lhcb experiment 
at $\sqs = 7 \tev$, corresponding to an integrated luminosity of 0.7\,fb$^{-1}$, 
and at $\sqs = 8 \tev$, corresponding to an integrated luminosity of 2.0\,fb$^{-1}$. 

Simulated samples of $\etac (1S)$ and \jpsi mesons decaying to the \proton\antiproton final state, 
and \jpsi decaying to the \proton\antiproton\piz final state, are used to estimate efficiency ratios, 
the contribution from the decay $\jpsi \to \proton\antiproton\piz$, 
and to evaluate systematic uncertainties. 
In the simulation, $pp$ collisions are generated using \pythia~\cite{Sjostrand:2006za} 
with a specific \lhcb configuration~\cite{LHCb-PROC-2010-056}.  
Decays of hadronic particles are described by \evtgen~\cite{Lange:2001uf}, 
in which final-state radiation is generated using \photos~\cite{Golonka:2005pn}. 
The interaction of the generated particles with the detector and its response are implemented 
using the \geant toolkit~\cite{Allison:2006ve, *Agostinelli:2002hh} as described in
Ref.~\cite{LHCb-PROC-2011-006}.

\section{Signal selection and data analysis}
\label{sec:ana}

The signal selection is largely performed at the trigger level. 
The offline analysis, in addition, requires the transverse momentum of \proton and \antiproton to be 
$\pt > 2.0 \gevc$, and restricts charmonium candidates to the rapidity range $2.0 < y < 4.5$. 

Discrimination between promptly produced charmonium candidates and those from 
$b$-hadron decays is achieved using the pseudo-decay time $t_{z} = \Delta z \cdot M / p_z$, 
where $\Delta z$ is the distance along the beam axis between 
the corresponding \proton\proton collision vertex (primary vertex) 
and the candidate decay vertex, $M$ is the candidate mass, and $p_z$ is the longitudinal component 
of its momentum. 
Candidates with $t_{z} < 80 \fs$ are classified as prompt, while those 
with $t_{z} > 80 \fs$ are classified as having originated from \bquark-hadron decays. 
For charmonium candidates from \bquark-hadron decays, a significant displacement 
of the proton tracks with respect to the primary vertex is also required. 

The selected samples of prompt charmonium candidates and charmonia from \bquark-hadron decays  
have some candidates wrongly classified 
(cross-feed). 
The cross-feed probability is estimated using simulated samples and is scaled using the observed 
signal candidate yields in data. 
The cross-feed component is subtracted to obtain the ratio of produced $\etac (1S)$ and \jpsi mesons 
decaying into the \proton\antiproton final state. 
Corrections range from 2\% to 3\% for the ratio of promptly produced $\etac (1S)$ and \jpsi mesons, 
and from 8\% to 10\% for the ratio of charmonia produced in \bquark-hadron decays. 

The ratios of signal yields are expressed in terms of 
ratios of cross-sections multiplied by the decay branching fractions as 
\begin{align*}
 \frac{N^{P}_{\etac (1S)}}{N^{P}_{\jpsi}} &= \frac{\sigma(\etac (1S)) \times \BF ( \etac (1S) \to \proton\antiproton )} 
 {\sigma(\jpsi) \times \BF ( \jpsi\to \proton\antiproton )}  , \\  
 \frac{N^{b}_{\etac (1S)}}{N^{b}_{\jpsi}} &= \frac{\BF ( \bquark \to \etac (1S) X ) \times \BF ( \etac (1S) \to \proton\antiproton )}
 {\BF ( \bquark \to \jpsi X ) \times \BF ( \jpsi \to \proton\antiproton )} ,
\end{align*}
where $N^P$ and $N^b$ are the numbers of charmonia 
from prompt production and \bquark-hadron decays, respectively. 
The simulation describes the kinematic-related differences between the $\etac (1S)$ and \jpsi decay modes reasonably well 
and predicts that the relative efficiencies for selecting and reconstructing $\etac (1S)$ and \jpsi mesons differ by less than 0.5\%.
Equal efficiencies are assumed for the $\etac (1S)$ and \jpsi meson reconstruction and selection criteria. 
The efficiency for selecting and reconstructing prompt \jpsi mesons 
is corrected for polarization effects, 
as a function of rapidity and \pt, according to Ref.~\cite{LHCb-PAPER-2013-008}. 

The numbers of reconstructed $\etac (1S)$ and \jpsi candidates are extracted 
from an extended maximum likelihood fit to the unbinned \proton\antiproton invariant mass distribution. 
The \jpsi peak position $M_{\jpsi}$ and the mass difference $\Delta M_{\jpsi , \etac (1S)}$ are fitted 
in the sample of charmonia from \bquark-hadron decays, 
where the signal is more prominent because of the reduced background level due to charmonium decay-vertex 
displacement requirements. 
The results are then used to apply Gaussian constraints in the fit 
to the \proton\antiproton invariant mass spectrum in the prompt production analysis, 
where the signal-to-background ratio is smaller, due to large combinatorial backgrounds. 

The signal shape is defined by the detector response, combined with the natural width in the case of the $\etac (1S)$ resonance. 
The detector response is described using two Gaussian functions with a common mean value. 
In the description of each resonance, 
the ratio of narrow to wide Gaussian widths, $\sigma^a_{\jpsi} / \sigma^b_{\jpsi} = \sigma^a_{\etac (1S)} / \sigma^b_{\etac (1S)}$, 
the fraction of the narrow Gaussian component, 
and the ratio of the $\etac (1S)$ and \jpsi narrow Gaussian widths, 
$\sigma^a_{\etac (1S)}/\sigma^a_{\jpsi}$, 
are fixed in the fit to the values observed in simulation. 
The only resolution parameter left free in the fit 
to the low-background sample from \bquark-hadron decays, $\sigma^a_{\jpsi}$, 
is fixed to its central value in the fit to the prompt sample. 
The natural width $\Gamma_{\etac (1S)}$ of the $\etac (1S)$ resonance is also extracted 
from the fit to the \bquark-hadron decays sample, 
and is fixed to that value in the prompt production analysis. 
Gaussian constraints on the \jpsi meson mass and the $\Delta M_{\jpsi , \, \etac (1S)}$ mass difference 
from the fit to the \bquark-hadron decays sample are applied in the prompt production analysis. 
The fit with free mass values gives consistent results. 

The combinatorial background is parametrized by an exponential function 
in the fit of the sample 
from \bquark-hadron decays, 
and by a third-order polynomial in the fit to the prompt sample. 

Combinations of \proton\antiproton from the decay $\jpsi \to \proton\antiproton \piz$ 
potentially affect the region close to the $\etac (1S)$ signal; 
hence, this contribution is specifically included in the background description. 
It produces a non-peaking contribution, 
and its mass distribution is described by a square-root shape to account for the phase space available 
to the \proton\antiproton system from the $\jpsi \to \proton\antiproton \piz$ decay, 
convolved with two Gaussian functions to account for the detector mass resolution. 
In the fit to the \proton\antiproton invariant mass spectrum, 
the normalization of this contribution is fixed 
using the number of candidates found in the \jpsi signal peak 
and the ratios of branching fractions and efficiencies 
for the $\jpsi \to \proton\antiproton \piz$ and $\jpsi \to \proton\antiproton$ decay modes. 

\begin{figure}[t]
\begin{center}
      \includegraphics[width=0.9\textwidth]{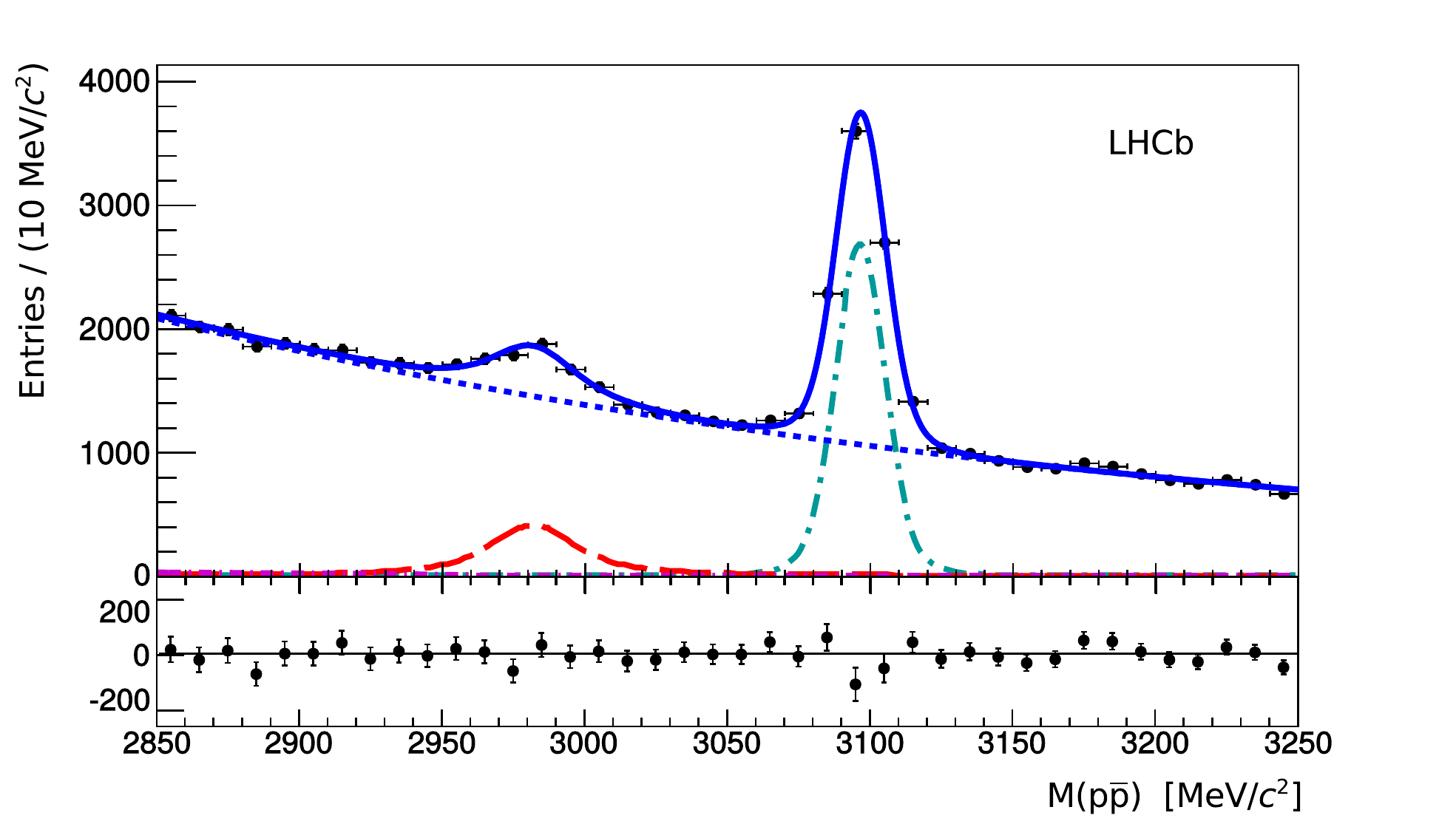}
\vspace*{-0.8cm}
\end{center}
\caption{Proton-antiproton invariant mass spectrum for candidates originating from a secondary vertex
and reconstructed in $\protect\sqs = 7 \protect\tev$ and $\protect\sqs = 8 \protect\tev$ data. 
The solid blue line represents the best-fit curve, 
the long-dashed red line corresponds to 
the $\protect\etac (1S)$ signal, 
the dashed-dotted cyan line corresponds to 
the \protect\jpsi signal, 
and the dashed magenta line corresponds 
to the small contribution from $\protect\jpsi \protect\to \protect\proton \protect\antiproton \protect\piz$ decays 
with the pion unreconstructed. 
The dotted blue line corresponds to the combinatorial background. 
The distribution of the difference between data points and the fit function is shown in the bottom panel.} 
\label{fig:seccombined}
\end{figure}

The \proton\antiproton invariant mass spectra for charmonium candidates from \bquark-hadron decays 
in the 7\tev and 8\tev data are observed to be 
consistent. 
The two data samples are therefore combined and the resulting 
spectrum is shown in Fig.~\ref{fig:seccombined} with the fit overlaid. 
The \jpsi meson signal is modelled using a double-Gaussian function. 
The $\etac (1S)$ signal is modelled using a relativistic Breit-Wigner function
convolved with a double-Gaussian function. 
The background contribution from the $\jpsi \to \proton\antiproton\piz$ decay with an unreconstructed pion, is small. 
The fit yields $2020 \pm 230$ $\etac (1S)$ signal decays and $6110 \pm 116$ \jpsi signal decays. 

\begin{figure}[t]
\begin{center}
		\includegraphics[width=0.9\linewidth]{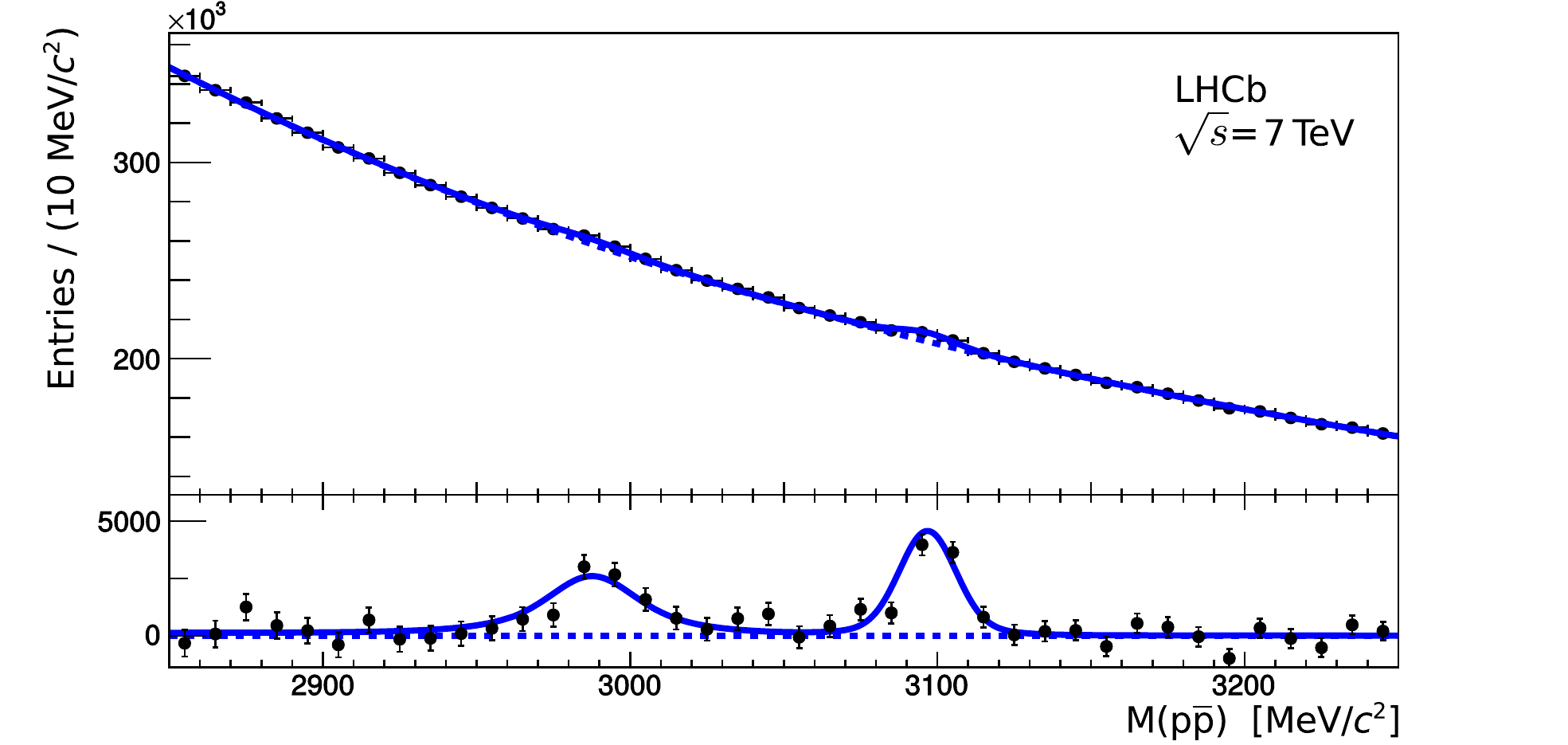}
                \put(-350,115) {\sffamily a)}
\vspace*{-0.6cm}
\end{center}
\begin{center}
		\includegraphics[width=0.9\linewidth]{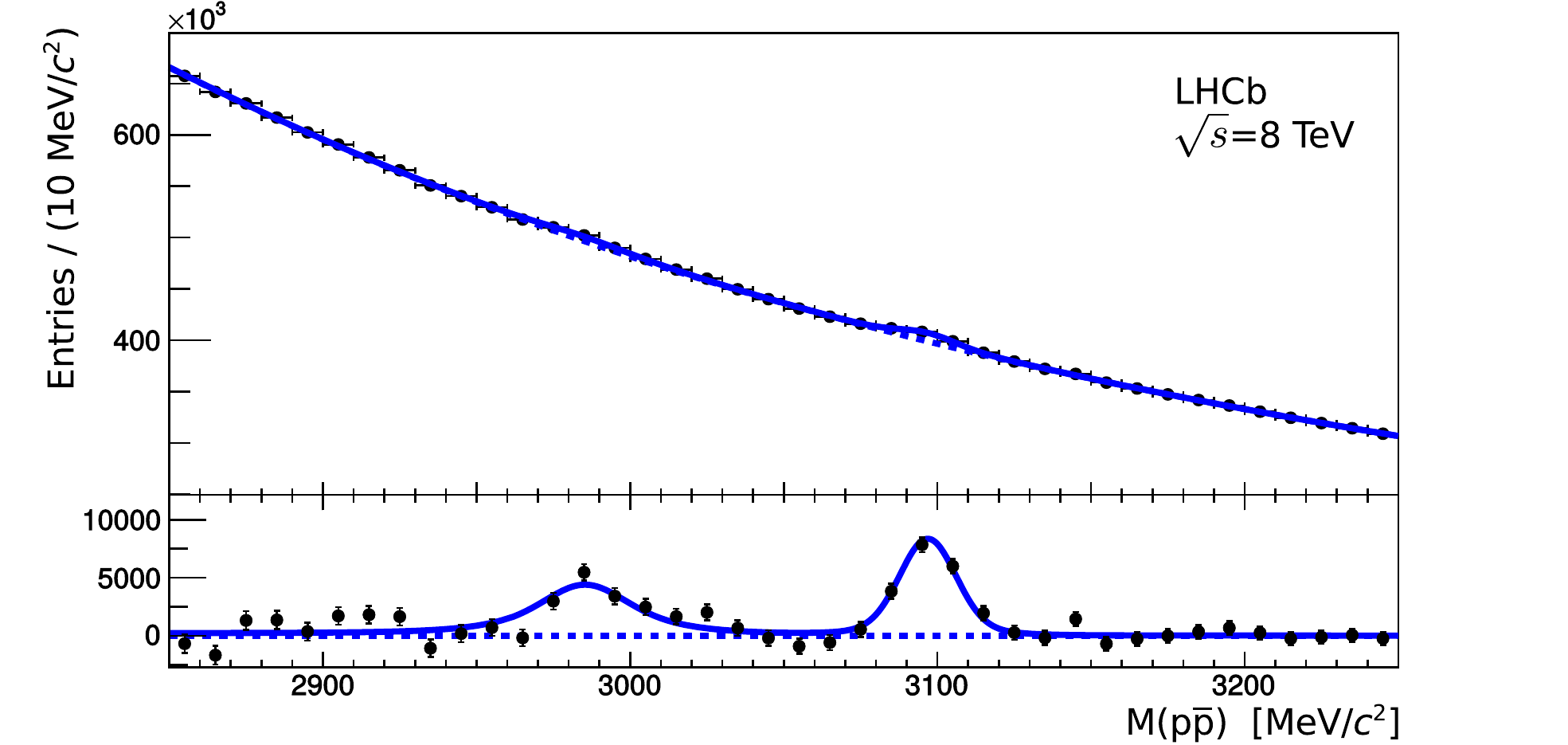}
                \put(-347,120) {\sffamily b)}
\vspace*{-0.8cm}
\end{center}
        \caption{Proton-antiproton invariant mass spectrum for candidates originating from a primary vertex 
(upper panel in each plot), 
and distribution of differences between data and the background distribution resulting from the fit 
(lower panel in each plot), 
in data 
at a) $\protect\sqs = 7 \protect\tev$ and b) $\protect\sqs = 8 \protect\tev$ centre-of-mass energies. 
Distributions on the upper panels are zero-suppressed.} 
\label{fig:pmtb}
\end{figure}

The results of the fit to the \proton\antiproton invariant mass spectrum of the prompt sample 
are shown in Fig.~\ref{fig:pmtb}a and~\ref{fig:pmtb}b 
for data collected at 
$\sqs = 7\tev$ and $\sqs = 8\tev$, respectively. 
The fits yield $13 \, 370 \pm 2260$ $\etac (1S)$ and $11 \, 052 \pm 1004$ \jpsi signal decays for the data taken at $\sqs = 7\tev$, 
and $22 \, 416 \pm 4072$ $\etac (1S)$ and $20 \, 217 \pm 1403$ \jpsi signal decays for the $\sqs = 8\tev$ data. 

In order to assess the quality of these unbinned fits to the invariant
$\proton \antiproton$ mass spectra, the chisquare per degree of freedom was calculated for
the binning schemes shown in Fig.~\ref{fig:seccombined}, \ref{fig:pmtb}a and~\ref{fig:pmtb}b. 
The values are 1.3, 1.7 and 1.8, respectively.

From the observed $\etac (1S)$ and \jpsi yields, and taking into account cross-feed between the samples, 
the yield ratios are obtained as 
\begin{align*}
\left( N^{P}_{\etac (1S)}/N^{P}_{\jpsi} \right)_{\sqs = 7\tev} &= 1.24 \pm 0.21 , \\ 
\left( N^{P}_{\etac (1S)}/N^{P}_{\jpsi} \right)_{\sqs = 8\tev} &= 1.14 \pm 0.21
\end{align*}
and 
\[
N^{b}_{\etac (1S)}/N^{b}_{\jpsi} = 0.302 \pm 0.039
\]
for the prompt production and charmonium production in \bquark-hadron decays. 
Only statistical uncertainties are given in the above ratios.

\section{Systematic uncertainties}
\label{sec:syst}

We consider systematic uncertainties due to limited knowledge of the detector mass resolution, 
the \jpsi polarization, 
the $\etac (1S)$ natural width, 
possible differences of the prompt charmonium production spectra in data and simulation, 
cross-feed between the prompt charmonium sample and the charmonium sample from \bquark-hadron decays, 
background description and feed-down from $\jpsi \to \proton \antiproton \piz$ decays. 
 \begin{table}[t] 
 \caption{Summary of uncertainties for the yield ratio $N_{\protect\etac (1S)} / N_{\protect\jpsi}$.}
{\small
\begin{center}\begin{tabular}{{l}*{3}{c}} 
			        & Production in  	& \multicolumn{2}{c}{Prompt production}   \\ 
			        & \bquark-hadron decays  	& $\sqs = 7\tev$        & $\sqs = 8\tev$  \\ \hline 
Statistical uncertainty 	
	 	                & $0.039$  	        & $0.21$   		& $0.21$   	\\ \hline 
Systematic uncertainties 
                                & & & \\ 
\hspace*{1.cm} Signal resolution ratio (simulation) 	
		                & $0.006$ 	        & $0.04$   		& $0.03$   	\\ 
\hspace*{1.cm} Signal resolution variation 	
	  	                &                       & $0.01$   		& $0.01$   	\\ 
\hspace*{1.cm} \jpsi polarization
	 			& $0.009$		& $0.02$		& $0.02$   \\ 
\hspace*{1.cm} $\Gamma_{\etac (1S)}$ variation 
	  	                & 	                & $0.15$   		& $0.14$   	\\ 
\hspace*{1.cm} Prompt production spectrum 	
	   			& $0.003$		& $0.07$   		& $0.06$   	\\ 
\hspace*{1.cm} Cross-feed 
	 			& $0.008$		& $0.01$		& $0.01$   \\ 
\hspace*{1.cm} Background model 
	 	                & $0.011$ 	        & $0.09$   		& $0.09$   \\ \hline 
Total systematic uncertainty 
	 	                & $0.018$ 	        & $0.20$   		& $0.18$   \\ 
\end{tabular}\end{center} 
\label{tab:syst}
}
 \end{table}
Uncertainties due to limited knowledge of the detector mass resolution are estimated by 
assigning the same $\sigma^a$ value to the $\etac (1S)$ and \jpsi signal description 
for the \bquark-hadron sample, 
and by varying the $\sigma^a$ parameters in the prompt production analysis within their uncertainties. 
Uncertainties associated with the \jpsi polarization in the prompt production 
reflect those of the polarization measurement in Ref.~\cite{LHCb-PAPER-2013-008}. 
We evaluate a potential contribution from \jpsi polarization in \bquark-hadron decays 
using a \babar study~\cite{Aubert:2002hc} of the \jpsi polarization in inclusive decays of \B mesons. 
Simulations are used to estimate the effective polarization parameter for the \lhcb kinematic region where the \bquark-hadrons
have a high boost and the longitudinal polarization is significantly reduced. 
A conservative value for the polarization parameter of -0.2 is used to estimate the corresponding systematic uncertainty.
%
In order to estimate the systematic uncertainty associated with the $\etac (1S)$ natural width, 
which enters the results for the prompt production analysis, the world average $\Gamma_{\etac (1S)}$ 
value of $32.0 \mev$ from Ref.~\cite{PDG2014} is used. 
Possible differences of the prompt charmonium production spectra in data and simulation 
are estimated by correcting the efficiency derived from simulation according to the observed \pt distribution. 
The uncertainty related to the cross-feed is estimated by varying 
the signal yields in each sample according to their uncertainties. 
Uncertainties associated with the background description are estimated by using an alternative 
parametrization and varying the fit range. 
The uncertainty due to the contribution from the $\jpsi \to \proton \antiproton \piz$ decay 
is dominated by the modelling of the $\proton \antiproton$ invariant mass shape, 
and is estimated by using an alternative parametrization, which is linear instead of the square root. 
Possible systematic effect related to separation between prompt and \bquark-decays samples, 
was checked by varying the $t_z$ discriminant value from 80 to $120 \fs$. 
The results are found to be stable under variation of the value of the $t_z$ discriminant, 
and no related systematic uncertainty is assigned. 
Table~\ref{tab:syst} lists the systematic uncertainties for the production yield ratio. 
The total systematic uncertainty is estimated as the quadratic sum of the
uncertainties from the sources listed in Table~\ref{tab:syst} and, in the case of the
prompt production measurement, is dominated by the uncertainty associated
with the $\etac (1S)$ natural width. For the measurement with \bquark-hadron decays
the uncertainties associated with the background model, the \jpsi
polarization and the cross-feed provide significant contributions.

%
%
%

\section{Results}
\label{sec:results}

The yield ratio for charmonium production in \bquark-hadron decays is obtained as 
\[
N^{b}_{\etac (1S)}/N^{b}_{\jpsi} = 0.302 \pm 0.039 \pm 0.015 . 
\]
In all quoted results, 
the first uncertainty refers to the statistical contribution and the second to the systematic contribution. 
By correcting the yield ratio with the ratio of branching fractions
$\BF ( \jpsi \to \proton \antiproton ) / \BF ( \etac (1S) \to \proton \antiproton ) = 1.39 \pm 0.15$~\cite{PDG2014}, 
the ratio of the inclusive \bquark-hadron branching fractions into $\etac (1S)$ and \jpsi final states 
for charmonium transverse momentum $\pt > 6.5 \gevc$ is measured to be 
\[
\BF ( \bquark \to \etac (1S) X ) / \BF ( \bquark \to \jpsi X ) = 0.421 \pm 0.055 \pm 0.025 \pm 0.045_{\BF} , 
\]
where the third uncertainty is due to that on the $\jpsi \to \proton \antiproton$ 
and $\etac (1S) \to \proton \antiproton$ branching fractions~\cite{PDG2014}. 
Assuming that the $\pt > 6.5 \gevc$ requirement does not bias the distribution 
of charmonium momentum in the \bquark-hadron rest frame, and 
using the branching fraction of \bquark-hadron inclusive decays into \jpsi mesons from Ref.~\cite{PDG2014}, 
$\BR ( \bquark \to \jpsi X ) = ( 1.16 \pm 0.10)\%$, 
the inclusive branching fraction of $\etac (1S)$ from \bquark-hadron decays is derived as 
\[
\BR ( b \to \etac (1S) X ) = (4.88 \pm 0.64 \pm 0.29 \pm 0.67 _{\BR}) \times 10^{-3} , 
\]
where the third uncertainty component includes also the uncertainty on the \jpsi inclusive branching fraction from \bquark-hadron decays. 
This is the first measurement of the inclusive branching fraction of \bquark-hadrons to an $\etac (1S)$ meson. 
It is consistent with a previous 90\% confidence level upper limit restricted to \Bub and \Bzb decays, 
$\BF ( \Bub , \Bzb \to \etac (1S) X ) < 9 \times 10^{-3}$~\cite{Balest:1994jf}. 

The prompt production yield ratios at the different centre-of-mass energies are obtained as 
\begin{align*}
\left( N^{P}_{\etac (1S)}/N^{P}_{\jpsi} \right)_{\sqs = 7 \tev} &= 1.24 \pm 0.21 \pm 0.20 , \\ 
\left( N^{P}_{\etac (1S)}/N^{P}_{\jpsi} \right)_{\sqs = 8 \tev} &= 1.14 \pm 0.21 \pm 0.18 .
\end{align*}
After correcting with the ratio of branching fractions 
$\BF ( \jpsi \to \proton \antiproton ) / \BF ( \etac (1S) \to \proton \antiproton )$~\cite{PDG2014}, 
the relative $\etac (1S)$ to \jpsi prompt production rates in 
the kinematic regime $2.0 < y < 4.5$ and $\pt > 6.5 \gevc$ 
are found to be 
\[
\left( \sigma_{\etac (1S)} / \sigma_{\jpsi} \right)_{\sqs = 7 \tev} = 1.74 \pm 0.29 \pm 0.28 \pm 0.18 _{\BF} , 
\]
for the data sample collected at 
$\sqs = 7 \tev$, and 
\[
\left( \sigma_{\etac (1S)} / \sigma_{\jpsi} \right)_{\sqs = 8 \tev} = 1.60 \pm 0.29 \pm 0.25 \pm 0.17 _{\BF} , 
\]
for the data sample collected at 
$\sqs = 8 \tev$. 
The third contribution to the uncertainty 
is due to that on the $\jpsi \to \proton \antiproton$ and $\etac (1S) \to \proton \antiproton$ branching fractions. 

The absolute $\etac (1S)$ prompt cross-section is calculated using the \jpsi prompt cross-section 
measured in Refs.~\cite{LHCb-PAPER-2013-008} and~\cite{LHCb-PAPER-2013-016} 
and integrated in the kinematic range of the present analysis, 
$2.0 < y < 4.5$ and $\pt > 6.5 \gevc$. 
The corresponding \jpsi prompt cross-sections were determined to be 
$( \sigma_{\jpsi} )_{\sqs = 7 \tev} = 296.9 \pm 1.8 \pm 16.9 \nb$ for $\sqs = 7 \tev$~\cite{LHCb-PAPER-2013-008}, and
$( \sigma_{\jpsi} )_{\sqs = 8 \tev} = 371.4 \pm 1.4 \pm 27.1 \nb$ for $\sqs = 8 \tev$~\cite{LHCb-PAPER-2013-016}. 
The \jpsi meson is assumed to be produced unpolarized. 
The prompt $\etac (1S)$ cross-sections in this kinematic region are determined to be 
\[
\left( \sigma_{\etac (1S)} \right)_{\sqs = 7 \tev} 
	= 0.52 \pm 0.09 \pm 0.08 \pm 0.06 _{\sigma_{\jpsi} , \, \BF} \mathrm{~\upmu b} , 
\]
for $\sqs = 7 \tev$, and 
\[
\left( \sigma_{\etac (1S)} \right)_{\sqs = 8 \tev} 
	= 0.59 \pm 0.11 \pm 0.09 \pm 0.08 _{\sigma_{\jpsi} , \, \BF} \mathrm{~\upmu b} , 
\]
for $\sqs = 8 \tev$. 
Uncertainties associated with the $\jpsi \to \proton \antiproton$ and $\etac (1S) \to \proton \antiproton$ branching fractions, 
and with the \jpsi cross-section measurement, are combined into the last uncertainty component, 
dominated by the knowledge of the branching fractions. 
This is the first measurement of prompt $\etac (1S)$ production in \proton\proton collisions. 
The cross-section for the $\etac (1S)$ prompt production is in agreement with 
the colour-singlet leading order (LO) calculations, while 
the predicted cross-section exceeds the observed value by two orders of magnitude
when the colour-octet LO contribution is taken into account~\cite{Biswal:2010xk}. 
However, the NLO contribution is expected to significantly modify the LO result~\cite{Maltoni:2004hv}. 
Future measurements at the \lhc design energy of $\sqs = 14 \tev$ 
may allow a study of the energy dependence of the $\etac (1S)$ prompt production. 

\begin{figure}[tb]
\begin{center}
                \includegraphics[width=0.9\textwidth]{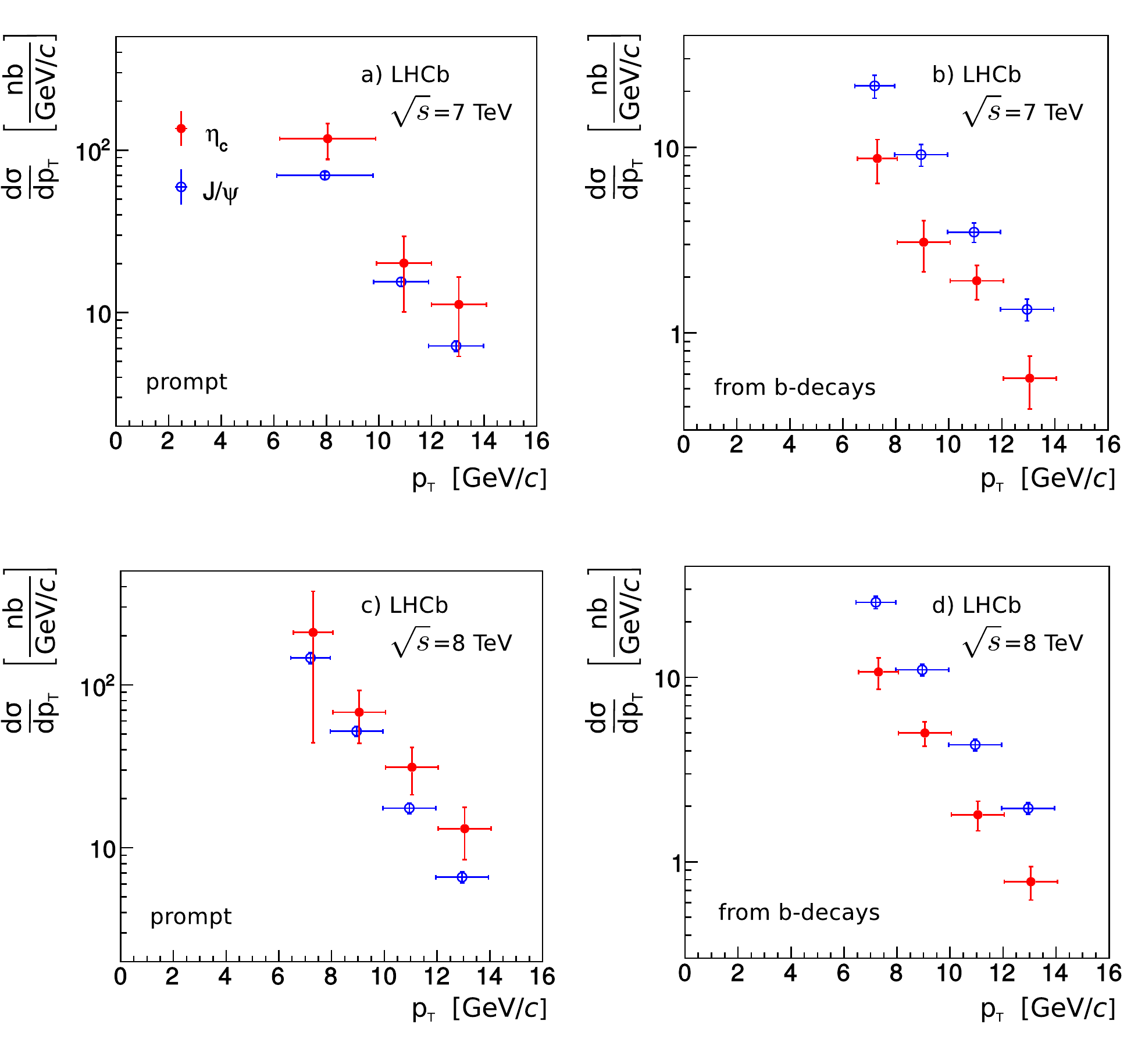}
\vspace*{-1.0cm}
 \end{center}
       \caption{Transverse momentum spectra for $\protect\etac (1S)$ mesons (red filled circles). 
          The \protect\pt spectra of \protect\jpsi from Refs.~\cite{LHCb-PAPER-2011-003,LHCb-PAPER-2013-008,LHCb-PAPER-2013-016} 
          are shown for comparison as blue open circles. 
	Prompt production spectra are shown on a) and c) for data 
	collected at $\protect\sqs = 7 \protect\tev$ and $\protect\sqs = 8 \protect\tev$, respectively. 
	The spectra from inclusive charmonium production in \bquark-hadron decays 
	are shown on b) and d) for data
	collected at $\protect\sqs = 7 \protect\tev$ and $\protect\sqs = 8 \protect\tev$, respectively.}
\label{fig:ptdata}
\end{figure}

The $\etac (1S)$ differential cross-section as a function of \pt is obtained by fitting the \proton\antiproton invariant mass spectrum 
in three or four bins of \pt. 
The same procedure as used to extract the $\etac (1S)$ cross-section is followed.
The \jpsi \pt spectrum measured in Refs.~\cite{LHCb-PAPER-2011-003,LHCb-PAPER-2013-008,LHCb-PAPER-2013-016} is used
to obtain the $\etac (1S)$ \pt spectrum for both prompt production 
and inclusive $\etac (1S)$ production in \bquark-hadron decays (Fig.~\ref{fig:ptdata}). 
The \pt dependence of the $\etac (1S)$ production rate exhibits similar behaviour to the \jpsi meson rate 
in the kinematic region studied. 

The performance of the \lhcb tracking system and the use of a final state common to \jpsi and $\etac (1S)$ decays 
allows a precise measurement of the mass difference between the two mesons.
In order to measure the $\etac (1S)$ mass relative to the well-reconstructed and well-known \jpsi mass, 
a momentum scale calibration~\cite{Aaij:2013qja} is applied on data, and validated with the \jpsi mass measurement. 
The $M_{\jpsi}$ and $\Delta M _{\jpsi , \, \etac (1S)}$ values are extracted from the fit to the \proton\antiproton invariant mass 
in the low-background sample of charmonium candidates produced in \bquark-hadron decays (Fig.~\ref{fig:seccombined}). 
The \jpsi mass measurement, $M_{\jpsi} = 3096.66 \pm 0.19 \pm 0.02$~\mevcc, 
agrees well with the average from Ref.~\cite{PDG2014}. 
The mass difference is measured to be 
\[
\Delta M_{\jpsi , \, \etac (1S)} = 114.7 \pm 1.5 \pm 0.1 \, \mevcc . 
\]
The systematic uncertainty is dominated by the parametrization of the $\jpsi \to \proton \antiproton \piz$ contribution. 
The mass difference agrees with the average from Ref.~\cite{PDG2014}. 
In addition, the $\etac (1S)$ natural width 
is obtained from the fit to the \proton\antiproton invariant mass (Fig.~\ref{fig:seccombined}), 
$\Gamma_{\etac (1S)} = 25.8 \pm 5.2 \pm 1.9 \mev$. 
The systematic uncertainty 
is dominated by knowledge of the detector mass resolution. 
The value of $\Gamma_{\etac (1S)}$ obtained is in good agreement with the average from Ref.~\cite{PDG2014}, 
but it is less precise than previous measurements.

\section{Summary}
\label{sec:summary}

In summary, $\etac (1S)$ production is studied using \proton\proton collision data corresponding 
to integrated luminosities of 0.7 fb$^{-1}$ and 2.0 fb$^{-1}$, 
collected at centre-of-mass energies $\sqs = 7 \tev$ and $\sqs = 8 \tev$, respectively. 
The inclusive branching fraction of \bquark-hadron decays into $\etac (1S)$ mesons with $\pt > 6.5 \gevc$, 
relative to the corresponding fraction into \jpsi mesons, 
is measured, for the first time, to be 
\[
\BF ( \bquark \to \etac (1S) X ) / \BF ( \bquark \to \jpsi X ) = 0.421 \pm 0.055 \pm 0.025 \pm 0.045 _{\BF}.
\]
The first uncertainty is statistical, the second is systematic, and the third is due to uncertainties 
in the branching fractions of $\etac (1S)$ and \jpsi meson decays to the \proton\antiproton final state. 
The inclusive branching fraction of \bquark-hadrons into $\etac (1S)$ mesons is derived as 
\[
\BR ( b \to \etac (1S) X ) = (4.88 \pm 0.64 \pm 0.29 \pm 0.67 _{\BR}) \times 10^{-3}, 
\]
where the third uncertainty component includes 
also the uncertainty on the inclusive branching fraction of \bquark-hadrons into \jpsi mesons. 
The $\etac (1S)$ prompt production cross-section in the kinematic region 
$2.0 < y < 4.5$ and $\pt > 6.5 \gevc$, 
relative to the corresponding \jpsi meson cross-section, is measured, for the first time, to be 
\begin{align*}
\left( \sigma_{\etac (1S)}/\sigma_{\jpsi} \right)_{\sqs = 7 \tev} 
	&= 1.74 \pm 0.29 \pm 0.28 \pm 0.18 _{\BF}, \\
\left( \sigma_{\etac (1S)}/\sigma_{\jpsi} \right)_{\sqs = 8 \tev} 
	&= 1.60 \pm 0.29 \pm 0.25 \pm 0.17 _{\BF}, 
\end{align*}
where the third uncertainty component is due to uncertainties 
in the branching fractions of $\etac (1S)$ and \jpsi meson decays to the \proton\antiproton final state. 
From these measurements, absolute $\etac (1S)$ prompt cross-sections are derived, yielding 
\begin{align*}
\left( \sigma_{\etac (1S)} \right)_{\sqs = 7 \tev} 
	&= 0.52 \pm 0.09 \pm 0.08 \pm 0.06 _{\sigma_{\jpsi} , \, \BF} \mathrm{~\upmu b}, \\
\left( \sigma_{\etac (1S)} \right)_{\sqs = 8 \tev} 
	&= 0.59 \pm 0.11 \pm 0.09 \pm 0.08 _{\sigma_{\jpsi} , \, \BF} \mathrm{~\upmu b}, 
\end{align*}
where the third uncertainty includes an additional contribution 
from the \jpsi meson cross-section. 
The above results assume that the \jpsi is unpolarized. 
The $\etac (1S)$ prompt cross-section is in agreement with the colour-singlet LO calculations, 
whereas the colour-octet LO contribution predicts a cross-section that exceeds the observed value by two orders of magnitude~\cite{Biswal:2010xk}. 
The \pt dependences of the $\etac (1S)$ and \jpsi production rates exhibit similar behaviour in the kinematic region studied. 
The difference between the \jpsi and $\etac (1S)$ meson masses is also measured, yielding 
$\Delta M_{\jpsi , \, \etac (1S)} = 114.7 \pm 1.5 \pm 0.1 \mevcc$, 
where the first uncertainty is statistical and the second is systematic. 
The result is consistent with the average from Ref.~\cite{PDG2014}.

\section*{Acknowledgements}

\noindent We would like to thank Emi Kou for motivating the studies of charmonium production 
in \lhcb using hadronic final states and the useful discussions regarding 
charmonium production mechanisms. 
We express our gratitude to our colleagues in the CERN
accelerator departments for the excellent performance of the LHC. We
thank the technical and administrative staff at the LHCb
institutes. We acknowledge support from CERN and from the national
agencies: CAPES, CNPq, FAPERJ and FINEP (Brazil); NSFC (China);
CNRS/IN2P3 (France); BMBF, DFG, HGF and MPG (Germany); SFI (Ireland); INFN (Italy); 
FOM and NWO (The Netherlands); MNiSW and NCN (Poland); MEN/IFA (Romania); 
MinES and FANO (Russia); MinECo (Spain); SNSF and SER (Switzerland); 
NASU (Ukraine); STFC (United Kingdom); NSF (USA).
The Tier1 computing centres are supported by IN2P3 (France), KIT and BMBF 
(Germany), INFN (Italy), NWO and SURF (The Netherlands), PIC (Spain), GridPP 
(United Kingdom).
We are indebted to the communities behind the multiple open 
source software packages on which we depend. We are also thankful for the 
computing resources and the access to software R\&D tools provided by Yandex LLC (Russia).
Individual groups or members have received support from 
EPLANET, Marie Sk\l{}odowska-Curie Actions and ERC (European Union), 
Conseil g\'{e}n\'{e}ral de Haute-Savoie, Labex ENIGMASS and OCEVU, 
R\'{e}gion Auvergne (France), RFBR (Russia), XuntaGal and GENCAT (Spain), Royal Society and Royal
Commission for the Exhibition of 1851 (United Kingdom).



\newpage
\addcontentsline{toc}{section}{References}
\setboolean{inbibliography}{true}
\bibliographystyle{LHCb}
\bibliography{main,LHCb-PAPER,LHCb-CONF,LHCb-DP}

\newpage
\centerline{\large\bf LHCb collaboration}
\begin{flushleft}
\small
R.~Aaij$^{41}$, 
C.~Abell\'{a}n~Beteta$^{40}$, 
B.~Adeva$^{37}$, 
M.~Adinolfi$^{46}$, 
A.~Affolder$^{52}$, 
Z.~Ajaltouni$^{5}$, 
S.~Akar$^{6}$, 
J.~Albrecht$^{9}$, 
F.~Alessio$^{38}$, 
M.~Alexander$^{51}$
% ...

{\footnotesize \it
% ...
$ ^{5}$Clermont Universit\'{e}, Universit\'{e} Blaise Pascal, CNRS/IN2P3, LPC, Clermont-Ferrand, France\\
$ ^{6}$CPPM, Aix-Marseille Universit\'{e}, CNRS/IN2P3, Marseille, France\\
% ...
$ ^{9}$Fakult\"{a}t Physik, Technische Universit\"{a}t Dortmund, Dortmund, Germany\\
% ...
$ ^{37}$Universidad de Santiago de Compostela, Santiago de Compostela, Spain\\
$ ^{38}$European Organization for Nuclear Research (CERN), Geneva, Switzerland\\
% ...
$ ^{40}$Physik-Institut, Universit\"{a}t Z\"{u}rich, Z\"{u}rich, Switzerland\\
$ ^{41}$Nikhef National Institute for Subatomic Physics, Amsterdam, The Netherlands\\
% ...
$ ^{46}$H.H. Wills Physics Laboratory, University of Bristol, Bristol, United Kingdom\\
% ...
$ ^{51}$School of Physics and Astronomy, University of Glasgow, Glasgow, United Kingdom\\
$ ^{52}$Oliver Lodge Laboratory, University of Liverpool, Liverpool, United Kingdom\\
% ...
}
\end{flushleft}

\end{document}